\documentclass[a4paper,11pt]{article}
\usepackage{pos}

\usepackage{amsmath}
\usepackage{sidecap}

\newcommand{\be}{\begin{equation}}
\newcommand{\ee}{\end{equation}}
\newcommand{\bea}{\begin{eqnarray}}
\newcommand{\eea}{\end{eqnarray}}

\def\CO{{\cal O}}

\title{A new framework to tune an improved relativistic heavy-quark action}
\ShortTitle{A new method to tune the RHQ action}

\author*[a]{D.~Giusti}
\author[a,b]{C.~Lehner}

\affiliation[a]{Universit\"at Regensburg, Fakult\"at f\"ur Physik,\\
  D-93040, Regensburg, Germany}

\affiliation[b]{Brookhaven National Laboratory,\\
  NY-11973, Upton, USA}

\emailAdd{davide.giusti@ur.de}

\abstract{We introduce a new non-perturbative method to tune the parameters of the Columbia formulation of an anisotropic, clover-improved relativistic heavy-quark (RHQ) action.
By making use of suitable observables which can be computed at a sequence of heavy-quark mass values, employing an $O(a)$-improved discretized action with domain-wall chiral fermion, and safely interpolated between the accessible heavy-quark mass region and the static point predicted by heavy-quark effective theory, we are able to precisely determine the unknown coefficients of the RHQ action.
In this proof-of-principle study we benefit from the RBC/UKQCD Iwasaki gauge configurations with $2+1$ flavors of dynamical quarks, at three values of the lattice spacing varying from $0.11$ to $0.062$ fm.
Preliminary results and applications to bottom spectroscopy are also presented.}

\FullConference{%
 The 38th International Symposium on Lattice Field Theory, LATTICE2021
  26th-30th July, 2021
  Zoom/Gather@Massachusetts Institute of Technology
}

\begin{document}
\maketitle

\section{Introduction}
\label{sec:intro}

Precise knowledge of mass spectrum, decay, and mixing properties of hadrons containing one or more bottom or charm quarks is essential to advancing our understanding of the parameters of the Standard Model.
Lattice QCD provides methods to compute these quantities from first principles.
However, lattice calculations with heavy quarks present special difficulties since in full QCD calculations, which properly include the effects of dynamical quarks, it is often impractical to use a sufficiently small lattice spacing to control the $O(ma)^n$ discretization errors directly.
This is especially true for the more compute-intensive lattice fermion formulations where the light quarks are treated in a fashion respecting chiral symmetry (such as domain-wall or overlap fermions).
These problems are addressed by using a number of improved heavy-quark actions designed to control or avoid these potentially important finite lattice spacing errors.

One way to address this challenge is to adopt the Fermilab or relativistic heavy-quark (RHQ) method~\cite{El-Khadra:1996wdx,Aoki:2001ra,Christ:2006us} in which extra axis-interchange asymmetric terms are added to the usual relativistic action.
This action can accurately describe heavy-quark systems provided the improvement coefficients it contains are properly adjusted. As the heavy-quark mass decreases, this action goes over smoothly to the ${\cal O}(a)$-improved fermion action of Sheikholeslami and Wohlert (SW)~\cite{Sheikholeslami:1985ij}.
The RHQ action applies for all values of the heavy-quark mass $m_{Q}$, for both heavy-heavy and heavy-light systems, and allows a continuum limit.
Once improved, it accurately describes energies and amplitudes of on-shell states containing heavy quarks whose spatial momentum $\vec p$ is small compared to the inverse lattice spacing.
It can be shown~\cite{Christ:2006us} that all errors of order $|\vec p a|$, $(m_{Q} a)^n$ and $|\vec p a| (m_{Q} a)^n$ for all non-negative integers $n$ can be removed if an anisotropic, clover-improved Wilson action is used for the heavy quark.
This action depends on three relevant parameters: the bare quark mass $m_0$, an anisotropy parameter $\zeta$ and the coefficient $c_P$ of an isotropic
Sheikholeslami and Wohlert term.

In order to exploit this RHQ approach, values for these three parameters are needed.
The bare charm or bottom quark mass, $m_0$, is determined from experiment, usually by equating the known mass of a physical state containing one or two heavy quarks with the mass determined from a lattice calculation with the RHQ action.
The remaining two parameters, $\zeta$ and $c_P$, may be estimated from lattice perturbation theory or determined with a non-perturbative technique.
The authors of Ref.~\cite{RBC:2012pds} have proposed to determine $\zeta$ and $c_P$ non-perturbatively by imposing two simple conditions.
The first condition is the often-exploited requirement that the energy of a specific heavy-heavy or heavy-light state depend on that state's spatial momentum in a fashion consistent with continuum relativity:  $E(\vec{p})^2 = \vec p^2 + M^2$.
The second constraint is that a specific mass splitting agree with its experimental value.

However, this approach has the disadvantage that a possible experimental prediction from lattice QCD, a non-trivial spin-spin splitting, cannot be made.
In this work, starting from the latter observation, we propose a new strategy to tune the parameters of the RHQ action allowing not to sacrifice lattice predictivity.
In particular, we compute the hyperfine mass splitting of a heavy-quark system at a sequence of $m_{Q}$ values using the ${\cal O}(a)$-improved action of domain-wall chiral fermions (DWF) and then we safely interpolate between the accessible heavy-quark mass region and the static point predicted by heavy-quark effective theory (HQET) to determine the value of the mass splitting.
That lattice determination is then used as an input to compute the RHQ parameters non-perturbatively.

As is described below, these three conditions yield quite precise results for the three unknown parameters.
This new tuning approach has the additional advantage that tuned RHQ parameters can be predicted for any heavy-hadron mass.
In the final section of this manuscript results and applications to bottom spectroscopy are also presented.
These determinations can be viewed as tests of QCD and can be used to explore the accuracy and limitations of the RHQ approach.
Preliminary results show that an extension of the new method to non-perturbative renormalization and operator improvement in a position-space scheme is possible, thus paving the way to the computation of phenomenologically-important charm and bottom decay constants and mixing matrix elements, which are needed for determinations of CKM matrix elements and constraints on the CKM unitarity triangle.
Dedicated investigations along those lines and preliminary applications to radiative leptonic decays of heavy pseudoscalar mesons~\cite{Kane:2019jtj,Kane:2021zee} are currently underway.

\section{Heavy-quark action}
\label{sec:RHQ}

The RHQ method provides a consistent framework for describing both light quarks ($am_0 \ll 1$) and heavy quarks ($am_0 \approx 1$)~\cite{El-Khadra:1996wdx,Aoki:2001ra,Christ:2006us}.
This approach relies upon the fact that, in the rest frame of bound states containing one or more heavy quarks, the spatial momentum carried by each heavy quark is smaller than the mass of the heavy quark: for heavy-heavy systems $|\vec{p}| \sim \alpha_s m_0$ and for heavy-light systems $|\vec{p}| \sim \Lambda_\textrm{QCD}$.
Then one can perform the usual Symanzik expansion in powers of the spatial derivative $D_i$ (which brings down powers of $a\vec{p})$.
Terms of all orders in the mass $m_0 a$ and the temporal derivative $D_0$ must however be included.
Thus, a suitable lattice formulation for heavy quarks should break the axis-interchange symmetry between the spatial and temporal directions.  

In this work we adopt the same anisotropic clover-improved Wilson action for heavy quarks as in~\cite{RBC:2012pds}:
\begin{align}
S_\textrm{lat} &= a^4 \sum_{x,x'} \bar{\psi}(x') \left( m_0 + \gamma_0 D_0 + \zeta \vec{\gamma} \cdot \vec{D} - \frac{a}{2} (D^0)^2 - \frac{a}{2} \zeta (\vec{D})^2+ \sum_{\mu,\nu} \frac{ia}{4} c_P \sigma_{\mu\nu} F_{\mu\nu} \right)_{x' x} \psi(x) \,,
\label{eq:HQAct}
\end{align}
where
\begin{align}
D_\mu\psi(x) &= \frac{1}{2a} \left[ U_\mu(x)\psi(x+\hat{\mu}) - U_\mu^\dagger(x-\hat{\mu})\psi(x-\hat{\mu}) \right] \,, \\
D^2_\mu \psi(x) &= \frac{1}{a^2} \left[  U_\mu(x) \psi(x+\hat{\mu}) + U_\mu^\dagger(x - \hat{\mu})\psi(x-\hat{\mu}) - 2 \psi(x)  \right] \,, \\
F_{\mu\nu} \psi(x)&= \frac{1}{8 a^2} \sum_{s,s'= \pm 1} s s' \left[ U_{s\mu}(x) U_{s'\nu}(x+s\hat{\mu}) U_{s\mu}^\dagger(x + s'\hat{\nu}) U_{s'\nu}^\dagger (x)- \textrm{h.c.} \right] \psi(x) \,,
\end{align}
and  $\gamma_\mu = \gamma_\mu^\dagger$ , $\{\gamma_\mu,\gamma_\nu\} = 2 \delta_{\mu\nu}$ and $\sigma_{\mu\nu} = \frac{i}{2} [ \gamma_\mu , \gamma_\nu ]$.
Christ, Li, and Lin showed in Ref.~\cite{Christ:2006us} that one can absorb all positive powers of the temporal derivative by allowing the coefficients $c_P$ and $\zeta$ to be functions of the bare-quark mass $m_0 a$.
Thus, by suitably tuning the three coefficients in the action -- the bare-quark mass $m_0 a$, anisotropy parameter $\zeta$, and clover coefficient $c_P$ -- one can eliminate errors of $\CO(|\vec{p}|a)$, $\CO([m_0a]^n)$, and $\CO(|\vec{pa}|[m_0a]^n)$ from on-shell Green's functions.
The resulting action can be used to describe heavy quarks with $m_0a \approx 1$ with discretization errors that are comparable to those for light-quark systems.  

\section{Simulation details}
\label{sec:sim}

The parameters of the RHQ action suitable for describing heavy quarks depend upon the choice of actions for the gauge fields and sea quarks.
In this work we perform our numerical lattice computations on the $N_f = 2+1$ flavor domain-wall fermion ensembles generated by the RBC and UKQCD Collaborations~\cite{RBC-UKQCD:2008mhs,RBC:2010qam,RBC:2014ntl}.
These lattices include the effects of three light dynamical quarks; the lighter two sea quarks are degenerate and we denote their mass by $m_l$, while the heavier sea quark, whose mass we denote by $m_h$, is a little heavier than the physical strange quark.
The RBC/UKQCD lattices combine the Iwasaki action for the gluons~\cite{Iwasaki:1983iya} with the five-dimensional domain-wall action for the fermions~\cite{Shamir:1993zy,Furman:1994ky}.

We make use of several ensembles with different light sea-quark masses and lattice volumes, at three values of the lattice spacing (from $a^{-1} = 1.785$ GeV to $a^{-1} = 3.148$ GeV).
Tab.~\ref{tab:lattices} summarizes the main parameters.

\begin{table*}[htb!]
\begin{center}
\begin{tabular}{cccccccc} \hline\hline
ensemble & $\left(L/a\right)^3 \times \left(T/a\right)$ & $L_s / a$ & $\approx a$(fm) & $am_l$ & $am_h$ & $\approx M_\pi$(MeV) & $N_{conf}$ \\ \hline
24I & $24^3 \times 64$ & $16$ &  0.11 &  0.005 & 0.04 & 340 & 37 \\
24Ih & $24^3 \times 64$ &  $16$ & 0.11 &  0.01 & 0.04 & 426 & 26 \\ \hline
32I(u) & $32^3 \times 64$ &  $16$ & 0.083 &  0.004 & 0.03 & 302 & 27 \\ \hline
32Ifine & $32^3 \times 64$ &  $12$ & 0.063 &  0.0047 & 0.0186 & 371 & 34 \\ \hline\hline
\end{tabular}
\end{center}
\caption{Lattice simulation parameters used in this study. The columns list the lattice volume, extra fifth-dimensional extent, approximate lattice spacing, light ($m_l$) and strange ($m_h$) sea-quark masses, approximate unitary pion mass and number of configurations.}
\label{tab:lattices}
\end{table*}

All ensembles use the Shamir action approximation to the sign function.
Light quarks are simulated at their unitary value $am_l^{sea} = am_l^{val}$ whilst the valence strange quark masses are tuned to their physical values as determined in Ref.~\cite{RBC:2014ntl} with the exception of the 32Iu ensemble where we simulate at the unitary strange quark mass.
All propagators are generated using $Z_2$-wall sources and all mode averaging (AMA)~\cite{Shintani:2014vja} with 6 sloppy and 1 exact samples per configuration is employed.

AMA helps us in improving the statistical precision of our correlation functions.
This is accomplished by computing correlators originating from many time slices spaced throughout the temporal extent of the lattice.
We compute reduced-precision (sloppy) correlation functions on all of the time slices and to compute full-precision (exact) correlation functions only on a subset of the time slices.
The difference between sloppy and exact solves on some time slices may be used as a correction for bias introduced by computing sloppy solves, and averages of sloppy solves on the remaining time slices improve statistical uncertainties at a reduced cost compared to computing exact solves for all time slices.

The ensembles listed in Tab.~\ref{tab:lattices} have already been used to study the light pseudoscalar meson sector; we can therefore take advantage of many results from this earlier work.
The amount of chiral symmetry breaking in the light-quark sector can be parameterized in terms of an additive shift to the bare domain-wall quark mass called the residual quark mass. For all our ensembles, the size of the residual quark mass is quite small (see Refs.~\cite{RBC:2010qam,RBC:2014ntl} for further details).

\section{Non-perturbative tuning of the RHQ action parameters}
\label{sec:method}

In this section we describe our new tuning approach to determine the RHQ parameters $\{m_0a, c_P, \zeta\}$.
We decide to study heavy-strange systems since both discretization errors and chiral extrapolation errors are expected to be small.

In particular, we start computing the hyperfine mass splitting $\Delta_{H_{Qs}} = M_{H^\star_{Qs}} - M_{H_{Qs}}$ of a heavy-strange meson at a sequence of heavy-quark mass values (up to $m_Q \approx m_b / 2$) using the DW action.
For each ensemble of Tab.~\ref{tab:lattices} we simulate a range of heavy-quark masses and compute the mass splittings between vector and pseudoscalar states.
Then, we interpolate between the available lattice data and the known static point of $\Delta_{H_{Qs}}$ predicted by HQET~\cite{Isgur:1989vq,Neubert:1992fk} to determine the mass splittings of heavy-strange systems.

We perform extrapolations to the continuum and infinite volume limits and to the physical pion point ($M_\pi^{phys} = 135$ MeV) adopting the following combined phenomenological ansatz
\bea
\Delta_{H_{Qs}} & = & \left( \frac{A_0}{M_{H_{Qs}}} + \frac{A_1}{M_{H_{Qs}}^2} \right) \left[ 1 + C^{(\pi)}_0 \delta_{M_\pi}  + C^{(\pi)}_1 \delta_{M_\pi}^2 + C^{(K)} \delta_{M_K} \right] \nonumber \\
& \cdot & \left[ 1 + D_0 a^2 + D_1 (am_Q)^2 + D_2 a^4 + F M_\pi^2 \frac{e^{-M_\pi L}}{(M_\pi L)^{3/2}}  \right]~,
\label{eq:DWF_hyp_fit}
\eea
where $\delta_{M_P} = \left( M_P - M_P^{phys}\right)$ for $P = \pi, K$ and $A_0, A_1, C^{(\pi)}_0, C^{(\pi)}_1, C^{(K)}, D_0, D_1, D_2, F$ are free parameters.
We estimate the main systematic effects by including/excluding different terms in the fit function (\ref{eq:DWF_hyp_fit}).
The sea-quark chiral extrapolation turns out to be the dominant systematic error, suggesting that a better control could be achieved by using RBC/UKQCD gauge ensembles close to the physical pion mass.

\begin{figure}[htb!]
\centering
\includegraphics[scale=0.25]{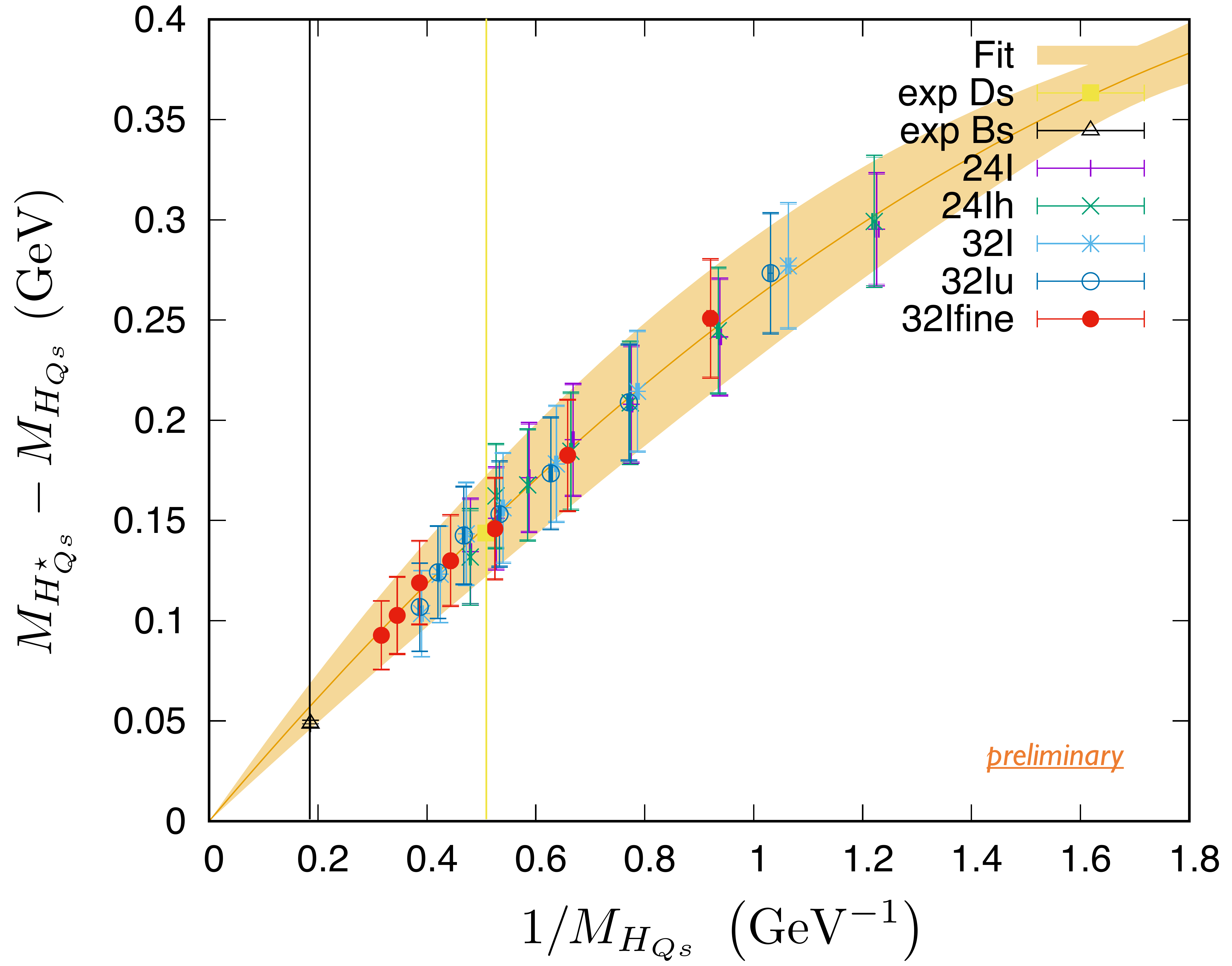}
\caption{Results for the hyperfine mass splitting $\Delta_{H_{Qs}} = M_{H_{Qs}^\star} - M_{H_{Qs}}$ of a heavy-strange meson $H_{Qs}$ versus $1/M_{H_{QS}}$ calculated for the DWF gauge ensembles of Tab.~\ref{tab:lattices}. For each ensemble a range of heavy-quark masses is simulated. The data points displayed are corrected for the fitted lattice artifacts of Eq.~(\ref{eq:DWF_hyp_fit}). The solid orange line represents our predictions for $\Delta_{H_{Qs}}$ extrapolated to the continuum and infinite volume limits and to the physical pion mass (i.e. by setting the $C$s and $D$s parameters of the fit ansatz (\ref{eq:DWF_hyp_fit}) to 0) and the orange area identifies the corresponding uncertainty at the level of one standard deviation. The black triangle and the yellow square correspond to the experimental determinations for the hyperfine splittings $\Delta_{B_s}$ and $\Delta_{D_s}$ \cite{Zyla:2020zbs}, respectively. The vertical black and yellow lines help to visualize the bottom and charm energy regions. Innermost error-bars correspond to the statistical error.}
\label{fig:DWF_hyp}
\end{figure}

Our results are shown in Fig.~\ref{fig:DWF_hyp}, where data points are corrected for the fitted lattice artifacts indicated in square brackets in (\ref{eq:DWF_hyp_fit}) and the fit band represents the predictions extrapolated to the continuum and infinite volume limits and to the physical pion mass.
We check {\it a posteriori} that our determinations for the hyperfine mass splittings both in the bottom and in the charm energy regions are consistent with the experimental measurements $M_{B_s^\star} - M_{B_s}$ and $M_{D_s^\star} - M_{D_s}$ \cite{Zyla:2020zbs} (black and yellow points of Fig.~\ref{fig:DWF_hyp}, respectively).

At the same time, we repeat the calculation of $\Delta_{H_{Qs}}$ at a sequence of $m_0a$ values using now the RHQ action in Eq.~(\ref{eq:HQAct}).
For each ensemble and heavy-quark mass, we use, as a first step, some educated guesses for the parameters $c_P$ and $\zeta$.

Because the lattice action breaks Lorentz symmetry, mesons receive corrections to their energy-momentum dispersion relation due to lattice artifacts
\begin{align} 
	(aE)^2 = (aM_1)^2 + \left( \frac{M_1}{M_2} \right) (a\vec{p})^2 + \CO( [a\vec{p}]^4) \,.
\label{eq:DispRel}
\end{align}
The quantities $M_1$ and $M_2$ are known as the rest mass and kinetic mass, respectively,
\begin{align}
	M_1 = E(\vec{p} =0) \,, \qquad M_2 = M_1 \times \left( \frac{\partial E^2}{\partial p_i^2} \right)^{-1}_{\vec{p} = 0} \,,
\end{align}
and will generally be different for generic values of the parameters $\{m_0a, c_P, \zeta\}$.

We require that the $H_{Qs}$ meson rest and kinetic masses are equal
\begin{align}
\frac{M_1^{H_{Qs}}}{M_2^{H_{Qs}}} = 1 \,,
\end{align}
so that the meson satisfies the continuum energy-momentum dispersion relation $E^2_{H_{Qs}}(\vec{p}) = \vec{p}^2_{H_{Qs}} + M^2_{H_{Qs}}$.

We determine the tuned values of the RHQ parameters non-perturbatively using an iterative procedure similar to the one adopted in \cite{RBC:2012pds}.
For a single step of the iteration procedure we compute the quantities $\left\{ M_{H_{Qs}}, M_{H_{Qs}^\star}, E_{H_{Qs}}(\vec{p} \neq 0) \right\}$ for each value of $m_0a$ at five sets of parameters in which we vary one of the two parameters $\{c_P, \zeta\}$ by a chosen uncertainty $\pm \sigma_{\{c_P, \zeta \}}$ while holding the other one fixed:
\bea
\left[\!\!\begin{array}{c} m_0 a\\ c_P\\ \zeta\\ \end{array}\right],
\left[\!\!\begin{array}{c}m_0 a\\c_P-\sigma_{c_P}\\ \zeta\\ \end{array}\!\!\right],\;
\left[\!\!\begin{array}{c}m_0 a\\c_P+\sigma_{c_P}\\ \zeta\\ \end{array}\!\!\right],\;
\left[\!\!\begin{array}{c}m_0 a\\ c_P\\ \zeta-\sigma_{\zeta}\\ \end{array}\!\!\right],\;
\left[\!\!\begin{array}{c}m_0 a\\ c_P\\ \zeta+\sigma_{\zeta}\\ \end{array}\!\!\right] \,.
\label{eq:FiveSets}
\eea
At fixed $m_0a$ value the heavy-strange meson masses in general will have a nonlinear dependence on the RHQ parameters $c_P$ and $\zeta$.
However, thanks to the ``box" of parameter space defined by the five parameter sets, we are able to test if we are working in a region sufficiently close to the true parameters to assume a linear approximation between the observables and the parameters. 
If indeed we are in the linear region, we then fit the five data sets for each quantity $X = \left\{ M_{H_{Qs}}, M_{H_{Qs}^\star}, E_{H_{Qs}}(\vec{p} \neq 0) \right\}$ using a simple linear function in $c_P$ and $\zeta$, i.e.
\be
X = a_X c_P + b_X \zeta + c_X ~ .
\label{eq:Fit_quantities}
\ee
Quadratic terms are used to estimate systematic errors.

Finally, the tuned RHQ parameters $\{c_P, \zeta\}$ for each value of $m_0a$ are obtained by solving the following system of equations
\begin{equation}
    \left[
      \begin{array}{c}
	\Delta_{H_{Qs}}\\
	\frac{M_1^{H_{Qs}}}{M_2^{H_{Qs}}}
      \end{array} \right]^{tuned} = J\cdot \left[
      \begin{array}{c}
        c_P \\
        \zeta
      \end{array}
      \right]^{\rm RHQ}
      + A \,,
      \label{eq:RHQDetermination}
\end{equation}
where $J$ and $A$ are a $2 \times 2$ matrix and a 2-element column vector, respectively, containing combinations of the known linear coefficients and constants $a_X, b_X$ and $c_X$ of Eq.~(\ref{eq:Fit_quantities}) and $\left[ M_1^{H_{Qs}} / M_2^{H_{Qs}} \right]^{tuned} = 1$.
In Eq.~(\ref{eq:RHQDetermination}) $\Delta_{H_{Qs}}^{tuned}$ is defined as
\be
\Delta_{H_{Qs}}^{tuned} = \frac{A_0}{M_{H_{Qs}}} + \frac{A_1}{M_{H_{Qs}}^2}
\ee
with $A_0$ and $A_1$ given by the DW fit (\ref{eq:DWF_hyp_fit}) and $M_{H_{Qs}}$ can be parameterized as in Eq.~(\ref{eq:Fit_quantities}).

We consider the RHQ parameters $\{c_P, \zeta\}$ to be tuned when both the values obtained via Eq.~(\ref{eq:RHQDetermination}) are within the ``box" defined by the five parameter sets in Eq.~(\ref{eq:FiveSets}).
This condition ensures that we are interpolating, rather than extrapolating, to the tuned point.
If the result for any of the parameters lies outside the box, we re-center the box around the result of Eq.~(\ref{eq:RHQDetermination}) and perform another iteration step.  We repeat this procedure until $\{c_P, \zeta\}$ tuned RHQ parameters lie inside the box for each $m_0a$.

\begin{figure}[htb!]
\centering
\includegraphics[scale=0.5]{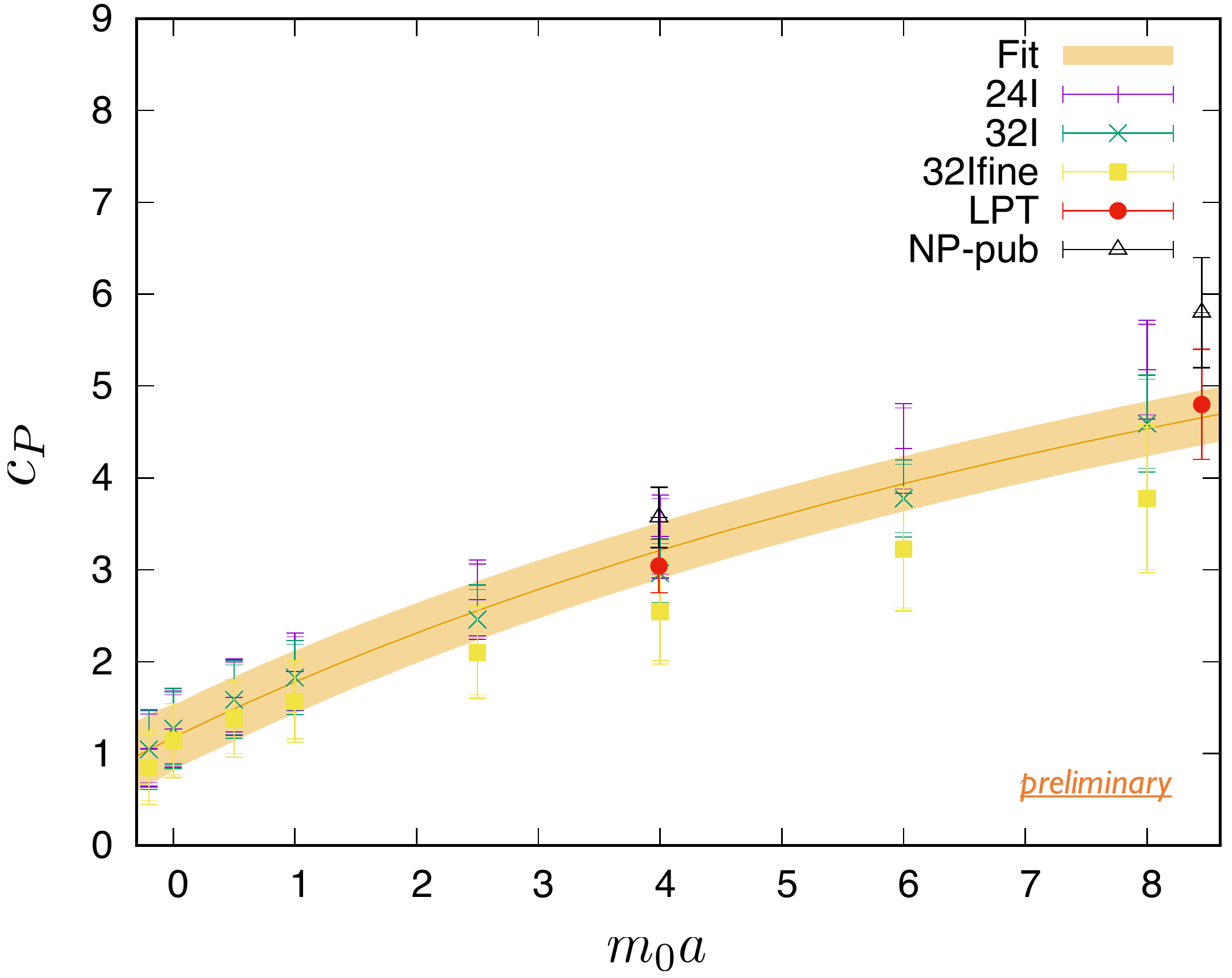}
\caption{Results for the tuned RHQ parameter $c_P$ as a function of the bare heavy-quark mass $m_0a$. Data points are corrected for the $\alpha_s$ and $a^2$ dependences included in the fit function (\ref{eq:RHQparams}). The orange band is the one sigma fit result with $\{ G_0, G_1, D \} = 0$. Black triangles represent previously published results for non-perturbative determinations of $c_P$~\cite{RBC:2012pds}, while red circles to the lattice perturbation theory estimates computed in Ref.~\cite{RBC:2012pds}. Inner error-bars correspond to statistical errors and outer error-bars to statistical and systematic uncertainties added in quadrature.}
\label{fig:cP}
\end{figure}

\begin{figure}[htb!]
\centering
\includegraphics[scale=0.5]{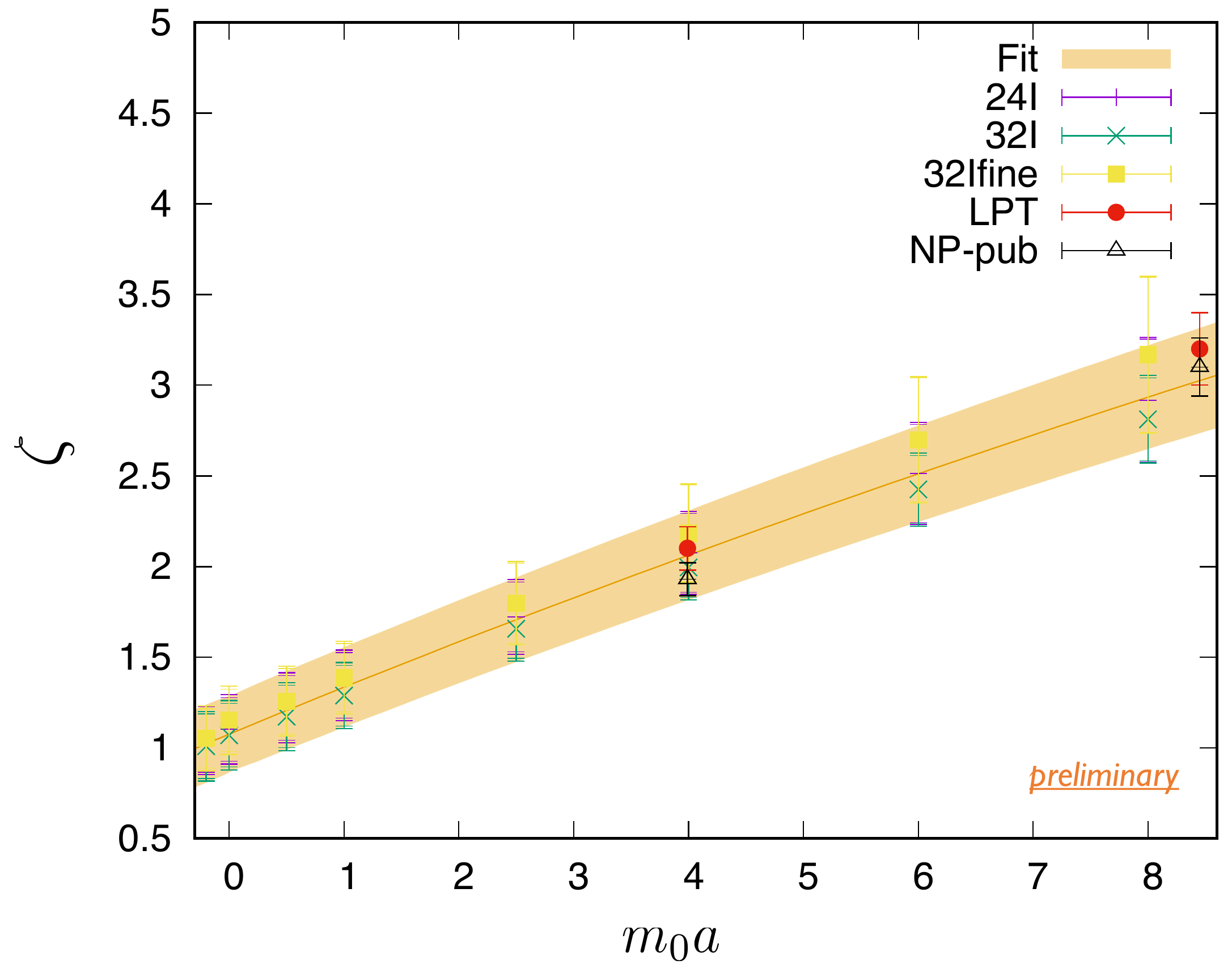}
\caption{The same of Fig.~\ref{fig:cP} for the tuned RHQ parameter $\zeta$.}
\label{fig:zeta}
\end{figure}

In Figs.~\ref{fig:cP} and \ref{fig:zeta} we show universality plots with our results for the tuned RHQ parameters $\{c_P, \zeta\}$ as a function of the bare heavy-quark mass $m_0a$.
We are able to predict tuned RHQ parameters in a wide range of the heavy-quark mass by fitting our lattice data to the following function inspired by perturbation theory
\be
\{c_P, \zeta\} = \frac{C_0 + C_1 m_0a}{C_2 + C_3 m_0a} \left[ 1+ G_0 \frac{\alpha_s(1/a)}{\pi} + G_1 \left(\frac{\alpha_s(1/a)}{\pi}\right)^2 \right] \left[1 + D a^2 \right]
\label{eq:RHQparams}
\ee
where $C_0, C_1, C_2, C_3, G_0, G_1, D$ are free parameters and the terms in square brackets parameterize the dependence of $\{c_P, \zeta\}$ on the QCD coupling constant.
A possible dependence of $\{c_P, \zeta\}$ on light-quark masses is expected to be weak and is therefore neglected in the present study.
Further refinements of the fit function (\ref{eq:RHQparams}) are postponed to future work.

In Figs.~\ref{fig:cP} and \ref{fig:zeta} the fit bands correspond to (\ref{eq:RHQparams}) with $\{ G_0, G_1, D \} = 0$ and the data points are corrected for the $\alpha_s$ and $a^2$ dependences.
Our non-perturbative predictions for $\{c_P, \zeta\}$ are in agreement both with previously published results and with lattice perturbation theory estimates~\cite{RBC:2012pds}.

Finally, as proposed in Ref.~\cite{RBC:2012pds}, $m_0a$ can be fixed by matching to the experimental values of the spin-averaged heavy-strange meson masses, $\overline{M}_{H_{Qs}} = \left(M_{H_{Qs}} + 3 M_{H_{Qs}^\star}\right)/4$.
We compute $\overline{M}_{H_{Qs}}$ at three sets of parameters in which we vary $m_0a$ by a chosen uncertainty and the other two parameters $\{c_P, \zeta\}$ are set at the values interpolated by (\ref{eq:RHQparams}).
The iterative procedure is repeated as described above.

\section{Applications to bottom spectroscopy}
\label{sec:results}

Once the RHQ parameters have been tuned, we can use them to predict physical on-shell quantities for heavy-light and heavy-heavy meson states.
We compute the desired quantities on the same sets of parameters used for the final iteration of the tuning procedure.
We then propagate the statistical errors in the tuned RHQ parameters to these quantities using the jackknife method;  this accounts for correlations between the RHQ parameters.

\begin{figure}[htb!]
\centering
\includegraphics[scale=0.5]{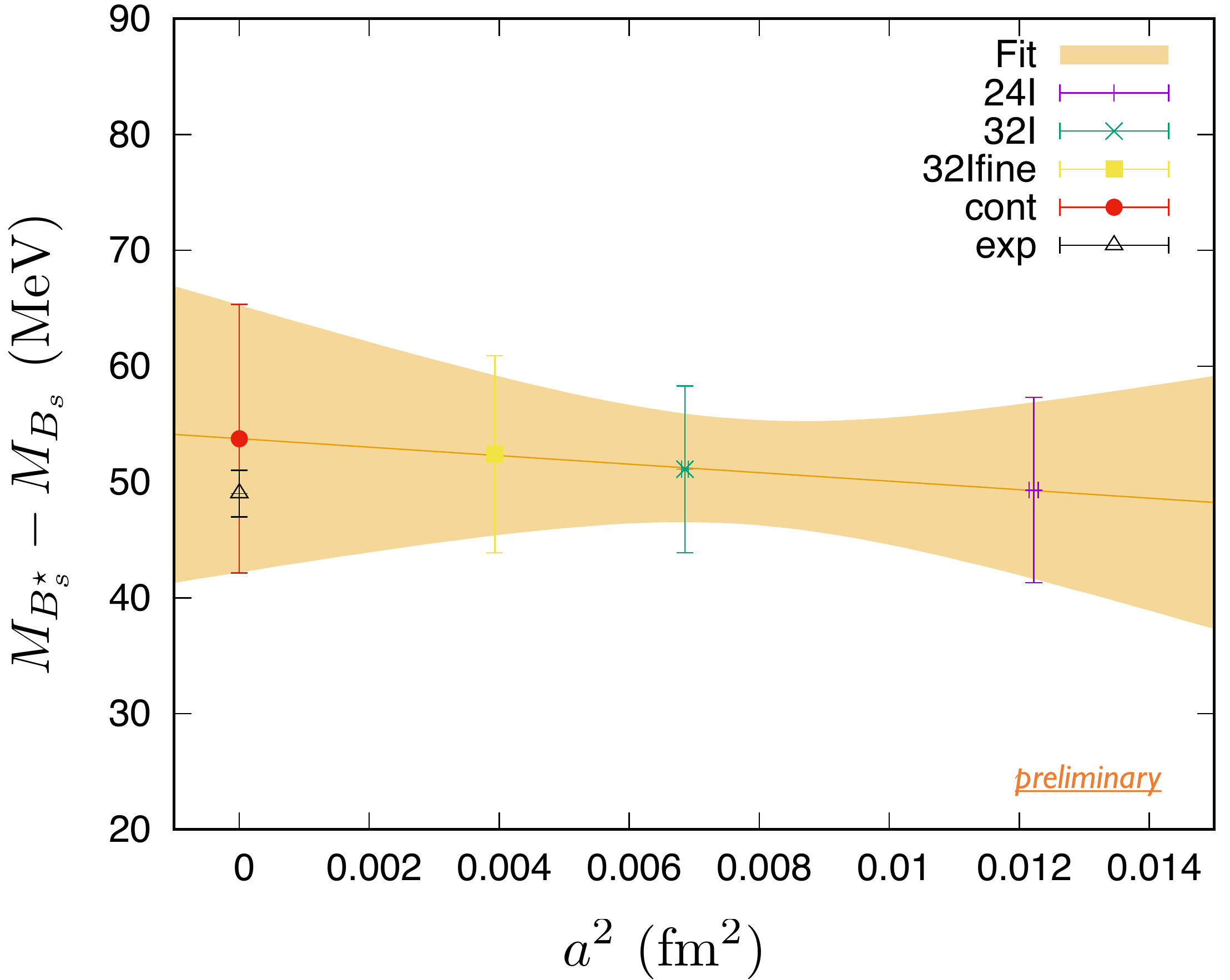}
\caption{Continuum extrapolation of the hyperfine mass splitting $M_{B_s^\star} - M_{B_s}$. Data points are already corrected for the chiral extrapolation in the light sea quark. The linear extrapolation in $a^2$ is indicated by the orange band and leads to the continuum limit result denoted by the red circle. For comparison we show the experimentally-measured value \cite{Zyla:2020zbs} as a black triangle.}
\label{fig:DBs}
\end{figure}

\begin{figure}[htb!]
\centering
\includegraphics[scale=0.5]{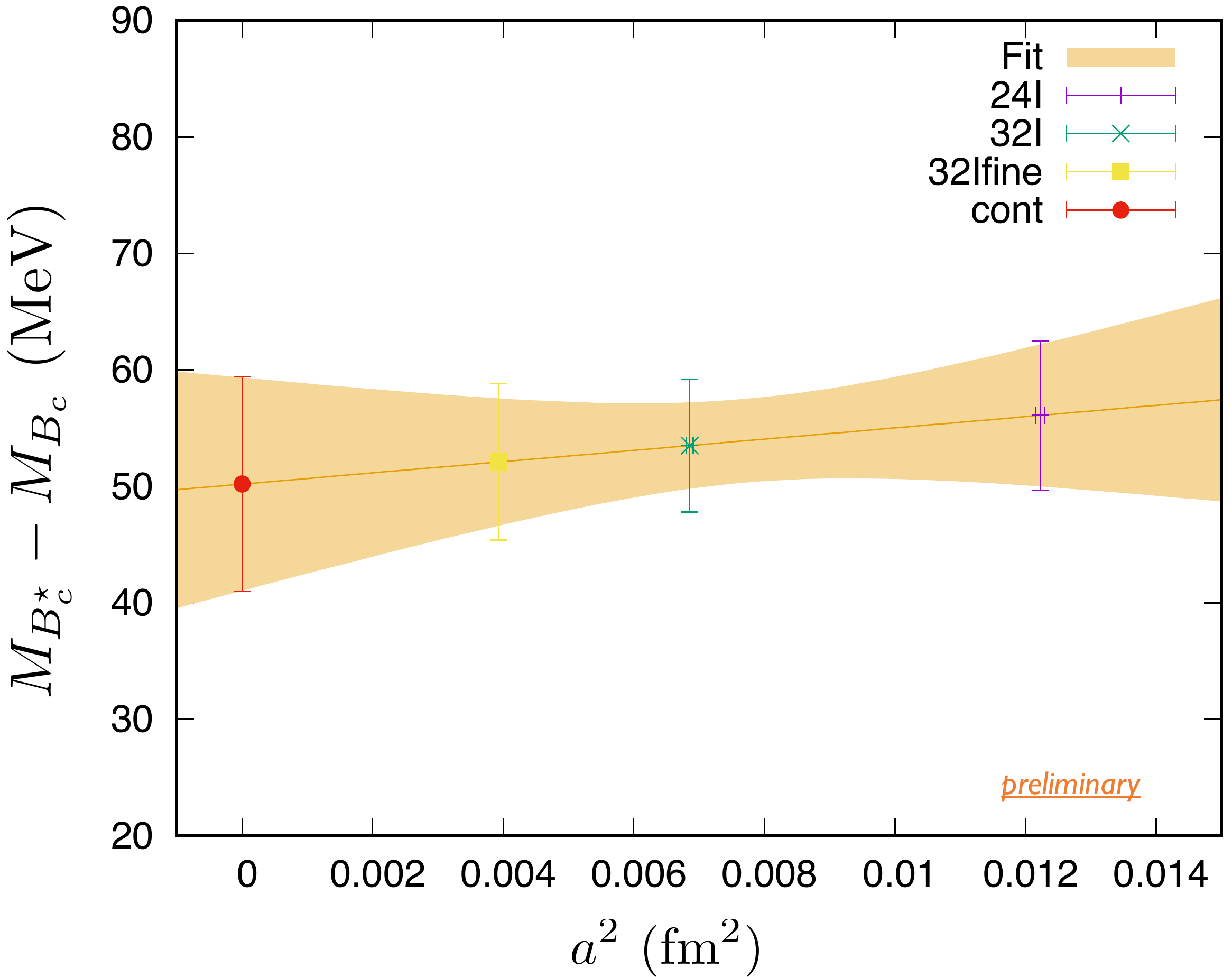}
\caption{The same as in Fig.~\ref{fig:DBs} in the case of the hyperfine mass splitting $M_{B_c^\star} - M_{B_c}$.}
\label{fig:DBc}
\end{figure}

In Figs.~\ref{fig:DBs} and \ref{fig:DBc} we show our preliminary determinations for the hyperfine mass splittings of the $B_s$ and $B_c$ systems on a subset of ${\cal O}(20)$ gauge configurations for the 24I, 32I and 32Ifine ensembles listed in Tab.~\ref{tab:lattices}.
Results are already extrapolated to the physical light-quark masses (we do not observe a statistically significant dependence of the observables on the light sea-quark masses).
Then, a linear extrapolation in $a^2$ leads to the following continuum limit results
\bea
\label{eq:DBs}
\label{eq:hypBsres}
M_{B_s^\star} - M_{B_s} & = & 54 ~ (11)_{stat} (3)_{syst} [11] ~ {\rm MeV} \\
\label{eq:hypBcres}
M_{B_c^\star} - M_{B_c} & = & 50 ~ (9)_{stat} (2)_{syst} [9] ~ {\rm MeV} ~,
\eea
where $()_{syst}$ represents the total systematic uncertainty with the various sources (discretization errors and continuum extrapolation, scale setting error, finite-volume effects, chiral extrapolation and systematic effects in the tuned RHQ parameters) added in quadrature.

Our preliminary lattice result (\ref{eq:DBs}) turns out to be consistent -- within our larger uncertainty -- with the experimental determination $\Delta_{B_s}^{exp} = 49 ~ (2) ~ {\rm MeV}$~\cite{Zyla:2020zbs}.
On the other hand an experimental measurements of the hyperfine mass splitting for the $B_c$ system is still lacking.
Thus, lattice QCD predictions provide new input which could be helpful for both experiments and phenomenological studies.
Our determination is in good agreement with the more precise lattice result $\Delta_{B_s} = 54 ~ (3) ~ {\rm MeV}$ of Ref.~\cite{Dowdall:2012ab}.
The improvement of the statistical uncertainties of (\ref{eq:hypBsres})-(\ref{eq:hypBcres}) is in progress.

\section*{Acknowledgments}
We thank the RBC and UKQCD Collaborations for providing the gauge-field configurations. 
We acknowledge PRACE for awarding us access to GCS@LRZ.


\begin{thebibliography}{99}

\bibitem{El-Khadra:1996wdx}
A.~X.~El-Khadra, A.~S.~Kronfeld and P.~B.~Mackenzie,
Phys. Rev. D \textbf{55}, 3933-3957 (1997)
[arXiv:hep-lat/9604004 [hep-lat]].

\bibitem{Aoki:2001ra}
S.~Aoki, Y.~Kuramashi and S.~i.~Tominaga,
Prog. Theor. Phys. \textbf{109}, 383-413 (2003)
[arXiv:hep-lat/0107009 [hep-lat]].

\bibitem{Christ:2006us}
N.~H.~Christ, M.~Li and H.~W.~Lin,
Phys. Rev. D \textbf{76}, 074505 (2007)
[arXiv:hep-lat/0608006 [hep-lat]].

\bibitem{Sheikholeslami:1985ij}
B.~Sheikholeslami and R.~Wohlert,
Nucl. Phys. B \textbf{259}, 572 (1985)

\bibitem{RBC:2012pds}
Y.~Aoki \textit{et al.} [RBC and UKQCD],
Phys. Rev. D \textbf{86}, 116003 (2012)
[arXiv:1206.2554 [hep-lat]].

\bibitem{Kane:2019jtj}
C.~Kane, C.~Lehner, S.~Meinel and A.~Soni,
PoS \textbf{LATTICE2019}, 134 (2019)
[arXiv:1907.00279 [hep-lat]].

\bibitem{Kane:2021zee}
C.~Kane, D.~Giusti, C.~Lehner, S.~Meinel and A.~Soni,
[arXiv:2110.13196 [hep-lat]].

\bibitem{RBC-UKQCD:2008mhs}
C.~Allton \textit{et al.} [RBC-UKQCD],
Phys. Rev. D \textbf{78}, 114509 (2008)
[arXiv:0804.0473 [hep-lat]].

\bibitem{RBC:2010qam}
Y.~Aoki \textit{et al.} [RBC and UKQCD],
Phys. Rev. D \textbf{83}, 074508 (2011)
[arXiv:1011.0892 [hep-lat]].

\bibitem{RBC:2014ntl}
T.~Blum \textit{et al.} [RBC and UKQCD],
Phys. Rev. D \textbf{93}, no.7, 074505 (2016)
[arXiv:1411.7017 [hep-lat]].

\bibitem{Iwasaki:1983iya}
Y.~Iwasaki,
[arXiv:1111.7054 [hep-lat]].

\bibitem{Shamir:1993zy}
Y.~Shamir,
Nucl. Phys. B \textbf{406}, 90-106 (1993)
[arXiv:hep-lat/9303005 [hep-lat]].

\bibitem{Furman:1994ky}
V.~Furman and Y.~Shamir,
Nucl. Phys. B \textbf{439}, 54-78 (1995)
[arXiv:hep-lat/9405004 [hep-lat]].

\bibitem{Shintani:2014vja}
E.~Shintani, R.~Arthur, T.~Blum, T.~Izubuchi, C.~Jung and C.~Lehner,
Phys. Rev. D \textbf{91}, no.11, 114511 (2015)
[arXiv:1402.0244 [hep-lat]].

\bibitem{Isgur:1989vq}
N.~Isgur and M.~B.~Wise,
Phys. Lett. B \textbf{232}, 113-117 (1989)

\bibitem{Neubert:1992fk}
M.~Neubert,
Phys. Rev. D \textbf{46}, 1076-1087 (1992)

\bibitem{Zyla:2020zbs}
P.A.~Zyla \textit{et al.} [Particle Data Group],
PTEP \textbf{2020}, no.8, 083C01 (2020)

\bibitem{Dowdall:2012ab}
R.~J.~Dowdall, C.~T.~H.~Davies, T.~C.~Hammant and R.~R.~Horgan,
Phys. Rev. D \textbf{86}, 094510 (2012)
[arXiv:1207.5149 [hep-lat]].

\end{thebibliography}
\end{document}